# Squeezed Josephson plasmons in driven YBa$_2$Cu$_3$O$_{6+x}$

N. Taherian[1], M. Först[1], A. Liu[1], M. Fechner[1], D. Pavicevic[1], A. von Hoegen[1], E. Rowe[1],

Y. Liu[2], S. Nakata[2], B. Keimer[2], E. Demler[3], M. H. Michael[1], A. Cavalleri[1,4]

[1] Max Planck Institute for the Structure and Dynamics of Matter, 22761 Hamburg, Germany
[2] Max Planck Institute for Solid State Research, 70569 Stuttgart, Germany
[3] Institute for Theoretical Physics, ETH Zurich, 8092 Zurich, Switzerland.
[4] Department of Physics, University of Oxford, Oxford OX1 3PU, United Kingdom

The physics of driven collective modes in quantum materials underpin a number of striking non-equilibrium functional responses, which include enhanced magnetism, ferroelectricity and superconductivity. However, the coherent coupling between multiple modes at once are difficult to capture by single-pump probe (one-dimensional) spectroscopy, and often remain poorly understood. One example is phonon-mediated amplification of Josephson plasmons in YBa$_2$Cu$_3$O$_{6+x}$, in which at least three normal modes of the solid are coherently mixed as a source of enhanced superconductivity. Here, we go beyond previous pump-probe experiments in this system and acquire two-dimensional frequency maps using pairs of mutually delayed, carrier envelope phase stable mid-infrared pump pulses, combined with measurements of the time-modulated second-order nonlinear optical susceptibility. We find that the driven zone-center phonons amplify coherent pairs of opposite-momentum Josephson plasma polaritons, generating a squeezed state of interlayer phase fluctuations. The squeezed state is a potentially important ingredient in the microscopic physics of photo-induced superconductivity in this and other materials.

Resonant optical driving of phonon modes has been shown to enhance superconductivity in a range of materials, including certain molecular solids and cuprate compounds [1-11]. In underdoped YBa$_2$Cu$_3$O$_{6+x}$, large-amplitude excitation of c-axis apical oxygen phonon modes was shown to induce transient optical features of non-equilibrium superconductivity above the transition temperature $T_C$ [2,6-8,12]. A representative example of this effect is shown in Figure 1a-c. Changes in the low-frequency complex optical constants were detected by electro-optically sampling c-axis-polarized, THz probe pulses after reflection from the sample (Fig. 1a, b). A near-2-THz plasma edge in the reflectivity was observed, as shown in Figure 1c[6]. This feature is associated with an imaginary optical conductivity that diverges like $1/\omega$ toward zero frequency (not shown), and has been discussed in terms of superconducting like c-axis transport.

Time- and angle-resolved second harmonic generation (SHG) measurements[8] acquired under these same excitation conditions have revealed that optically induced superconductivity is connected to the coherent dynamics of Josephson plasma polaritons (JPPs). JPPs are dispersive superconducting plasmons sustained by Cooper pair tunneling between the $CuO_2$ planes[13-17]. These modes are well formed below $T_C$ down to zero momentum. Above $T_C$, in the so-called pseudo-gap phase, the JPP modes may still be present, but are likely to be overdamped near zero momentum [18-20]. As shown in the representative plots of Fig. 1d-f, the coherent dynamics of these modes can be probed by time resolved second-harmonic generation. Because SHG vanishes in a centrosymmetric medium, inversion-symmetry odd (infrared-active) modes break mirror symmetry twice per cycle as they oscillate coherently. Hence, by measuring the SHG intensity as a function of pump probe time delay (sketched in Fig. 1d), and provided that the mid-infrared drive has a stable carrier-envelope phase, one obtains the response of all coherently oscillating infrared-active modes (see Fig. 1e). In the case of $YBa_2Cu_3O_{6+x}$, of which we show a representative Fourier amplitude spectrum in Fig. 1f, these modes include phonons and JPP oscillations. The two peaks at about 17 and 20 THz (shaded in yellow) reflect the resonantly driven infrared-active apical oxygen phonons, which are present also at small amplitudes of the drive. At the high drive fields, relevant for photo-induced superconductivity, one also finds oscillations near 2 THz, in the same frequency region where the Josephson plasma edge was observed in the THz reflectivity (Fig. 1c). The experimental results shown in Fig. 1d-f have been interpreted as the parametric amplification of finite-momentum JPPs, and connected to the mechanism of photo-induced superconductivity[8].

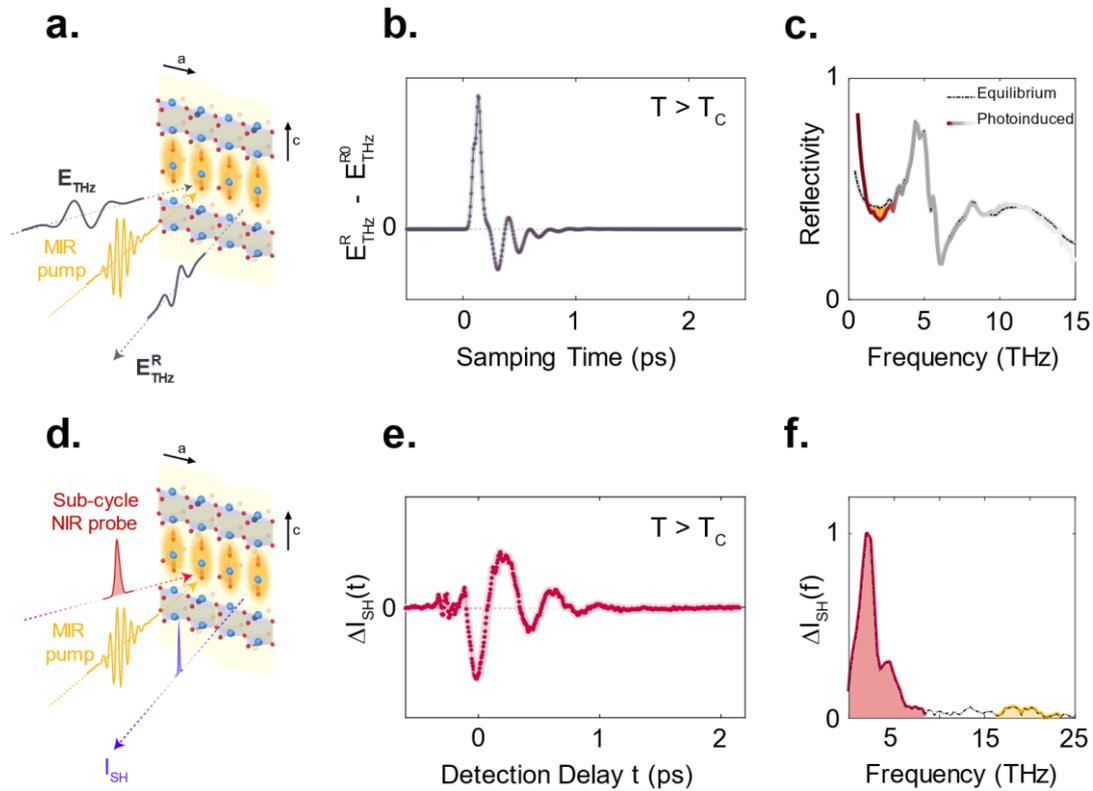

**FIG.1. (a)** Schematic of the mid-IR pump – THz probe experiment in YBa$_2$Cu$_3$O$_{6.48}$. The sample is excited by a mid-IR pump pulse (yellow) polarized along the crystal c-axis, resonantly driving apical oxygen phonon modes as indicated inside the yellow shading. The subsequent changes in the low-frequency optical properties are sampled by a broadband, also c-polarized THz probe pulse (grey). **(b)** Photo-induced change in the reflected THz probe electric field at the peak of mid-IR pump-THz probe response, measured at a base temperature of 60 K (Tc = 48 K) **(c)** Sample reflectivity in equilibrium (dashed gray line) and following photo-excitation (solid dark red line at low frequencies < 3 THz, solid grey line at frequencies above), measured at a base temperature of 60 K. The photo-induced changes are highlighted by yellow shading [6]. **(d)** Schematic of the mid-IR pump – time-resolved second harmonic probe experiment in the same sample. Here, the mid-IR pump pulse (yellow) is CEP stable and the subsequent dynamics are probed by collecting the second harmonic intensity (blue) generated from an 800 nm femtosecond probe pulse (red) as a function of detection delay t. Both incident pulses are polarized along the crystal c-axis. **(e)** Changes in the second harmonic intensity as a function of pump-probe time delay measured at T=100 K, above T$_c$. **(f)** Corresponding Fourier spectrum, highlighting the two resonantly excited apical oxygen phonons at 17 and 20 THz in yellow, and the low-frequency amplified Josephson plasmon in red [8].

In this paper, we start from the observation that the three-mode (one phonon and two plasmon) parametric mechanism discussed in the earlier work does not provide a unique explanation for these results. A four-wave mechanism, involving two phonons and two plasmons, is also consistent with these measurements. This ambiguity is understood based on the following considerations.

The measured time-delay-dependent modulations of the probe SHG intensity $\Delta I_{SH}$ for coherent oscillations of an infrared-active mode $Q_i(\omega_i)$ at frequency $\omega_i$ can be cast in the stimulated hyper-Raman scattering formalism. For a near-infrared probe field at 800 nm wavelength ($\omega_{pr} = 375\ THz$), interaction with a medium in which $Q_i(\omega_i)$ oscillates, causes radiation of an oscillatory second harmonic field arising from a hyper-Raman polarization $P_i(2\omega_{pr} \pm \omega_i) = \frac{\partial \chi^{(2)}}{\partial Q_i} Q_i(\omega_i) E_{pr}^2(\omega_{pr})$[21]. Provided that the pump pulse has a stable carrier envelope phase and that the probe pulses are shorter than one quarter of the oscillation period of the mode $Q_i(\omega_i)$, the radiated hyper-Raman field $E_i$ can be measured as pump-probe-delay dependent oscillations $\Delta I_{SH}$ of the SHG intensity. In absence of other electromagnetic fields on the detector, one finds oscillations proportional to $\Delta I_{SH,Hom} \sim |\sum_i E_i|^2$, which occur at twice the mode frequency $2\omega_i$ [22-25]. This channel is known as homodyne detection $\Delta I_{SH,Hom}$.

However, if interference with an auxiliary second-harmonic field takes place on the detector, time-delay-dependent coherent oscillations are found also at the frequency of the mode $\omega_i$. This detection scheme will be referred to as heterodyne detection $\Delta I_{SH,Het}$, whereby the auxiliary field $E_{LO}$ is known as a local oscillator and $\Delta I_{SH,Het} \sim \sum_i |E_i| |E_{LO}| \cos\phi$. For a sufficiently large field $E_{LO}$, the heterodyne channel can be made much larger than the homodyne signal.

The interpretation of Ref.[8] was based on the observation that a spurious, time delay independent second harmonic field $E_{LO}$ at 400 nm wavelength ($2\omega_{pr}$=750 THz), was co-propagating with the time-delay-dependent probe at the fundamental frequency $\omega_{pr}$ and with the second harmonic generated on the sample $E_i(2\omega_{pr} \pm \omega_i)$, serving as a local oscillator and yielding a response in the heterodyne channel at frequency $\omega_i$. A three-wave mixing process, sketched in Fig. 2a, was then proposed to explain the data. It was argued that the mid-infrared pulse resonantly excites the two apical oxygen phonons with coordinates $Q_{IR1}$ and $Q_{IR2}$ at frequencies $\omega_{IR1}/2\pi = 17$ THz and $\omega_{IR2}/2\pi = 20$ THz, respectively. Nonlinear mixing of *only one* of the two phonons ($Q_{IR1}$) with two JPPs (with current coordinates $J_{P1}$ and $J_{P2}$ corresponding to the inter-bilayer and intra-bilayer tunneling modes at frequencies $\omega_{JP1}$ and $\omega_{JP2}$ respectively), was considered as a source of parametric amplification of a pair of JPPs at finite momenta $\pm q_x$ along the in-plane direction *x*. This effect was described by a Hamiltonian interaction term $V_i = \alpha_1 q_x^2 Q_{IR1} J_{P2,q_x} J_{P2,-q_x}$. Due to simultaneous energy and momentum conservation, the frequency of the parametrically amplified JPP modes is constrained by the relation $\omega_{IR1} = \omega_{JP1}(-q_x) + \omega_{JP2}(q_x)$. A simulation of the time-delay-dependent phonon and plasmon displacements in this three-wave mixing scenario is shown in Fig. 2b. The calculated SHG spectra, assuming heterodyne detection, reproduce the experimentally measured ones, as shown in Fig. 2c. Two (yellow-shaded) peaks appear at the apical oxygen phonon frequencies of 17 and 20 THz. A large (red-shaded) peak around 2-3 THz and a weaker (also red-shaded) contribution at 13–14 THz were observed. These frequencies match those of the lower and upper JPPs observed below $T_C$ in equilibrium, and were measured at an in-plane wave vector $q_x$ of about 200 cm$^{-1}$ (see Ref.[8]).

However, alternative parametric coupling mechanisms may also explain the same observations. Because the mid-infrared excitation pulse simultaneously drives the phonon modes $Q_{IR1}$ and $Q_{IR2}$, a four-mode interaction involving the two phonons and a pair of low-frequency inter-bilayer JPPs $J_{P1,\pm q_x}$ cannot be excluded. This four-mode coupling would result from an interaction Hamiltonian $V_i = \beta(Q_{IR1} + Q_{IR2})^2 J_{P1,q_x} J_{P1,-q_x}$ (see Fig. 2d and Ref.[26]), which contains a resonance condition $2\omega_{JP1}(\pm q_x) = \omega_{IR2} - \omega_{IR1}$ for the mixing component $\beta(Q_{IR1}Q_{IR2})J_{P1,q_x}J_{P1,-q_x}$. Here, the coherent excitation of the two phonons leads to the parametric amplification of two lower-frequency JPPs, $J_{P1,q_x}$ and $J_{P1,-q_x}$, at frequency $\omega_{JP1}$ and identically-opposite momenta $\pm q_x$. In this four-mode mixing scenario, originally disregarded *in Ref.[8],* one would not expect to find signatures of $J_{P1,q_x}$ and $J_{P1,-q_x}$ in the heterodyne detection. Because the amplified plasmons $J_{P1,q_x}$ and $J_{P1,-q_x}$ start from a fluctuating seed with a temporal phase that changes from shot to shot, the overall response, when accumulated over many pulses in the experiment, could readily be expected to average to zero: $\langle J_{P1,q_x}\rangle = \langle J_{P1,-q_x}\rangle = 0$, hence $\Delta I_{SH,Het} = 0$. Shot-to-shot phase noise is subtracted away when considering the covariance of the two modes $\langle J_{P1,q_x} J_{P1,-q_x}\rangle$, which forms a squeezed state[27-29] that oscillates at $2\omega_{JP1}$ with a fixed phase for all the pulses of the experiment. However, this squeezed mode $\langle J_{P1,q_x} J_{P1,-q_x}\rangle$ is symmetry-even and not hyper-Raman active, hence it does neither break inversion symmetry nor modulate the SHG intensity in the heterodyne detection (see Supplementary Information for details). On this basis, the four-mode mechanism was excluded in the earlier work.

After more detailed inspection, the four-mode mixing mechanism should not be disregarded. It appears that the amplified pairs $J_{P1,q_x}$ and $J_{P1,-q_x}$, are bound to respond coherently by the parametric process. In fact, the difference-frequency component of the

$Q_{IR1}Q_{IR2}$ drive, which is at frequency $\omega_{IR2} - \omega_{IR1}$, has a constant absolute phase for all laser shots, even if the absolute phase of each phonon $Q_{IR1}$ or $Q_{IR2}$ was fluctuating from shot to shot. The phase of the parametrically amplified JPP pairs will then be set to the phase of the $Q_{IR1}Q_{IR2}$ difference-frequency component, plus a zero or a $\pi$-shift for each shot. Hence, even after averaging over many excitation pulses in the experiment, the amplified plasmon responses $\langle J_{P1,q_x}\rangle$ and $\langle J_{P1,-q_x}\rangle$ remain zero. Figure 2e shows the simulated dynamics of one amplified plasmon pair $J_{P1,q_x}$ and $J_{P1,-q_x}$ for one single laser shot, alongside the averaged responses $\langle J_{P1,\pm q_x}\rangle$.

On the other hand, each individually amplified plasmon $J_{P1,q_x}$ and $J_{P1,-q_x}$ is hyper-Raman active, hence each one of these two modes will generate a separate hyper-Raman polarization $P_{J_{P1,\pm q_x}}(2\omega_{pr.} \pm \omega_{J_{P1,\pm q_x}}) = \frac{\partial \chi^{(2)}}{\partial J_{P1,\pm q_x}} J_{P1,\pm q_x}(\omega_{J_{P1,\pm q_x}}) E_{pr.}^2(\omega_{pr.})$ in the material. Two separate fields $E_{J_{P1,+q_x}}(2\omega_{pr.} \pm \omega_{J_{P1,+q_x}})$ and $E_{J_{P1,-q_x}}(2\omega_{pr.} \pm \omega_{J_{P1,-q_x}})$ will be emitted. These generate a homodyned SHG contributions $\Delta I_{SH,Hom} \sim |E_{J_{P1,+q_x}} + E_{J_{P1,-q_x}}|^2$ which yields a modulation at the frequency of the squeezed plasmon oscillations at $2\omega_{JP1}$ via their mixing term $\Delta I_{SH,Hom} \sim E_{J_{P1,+q_x}} E_{J_{P1,-q_x}}$ whilst their heterodyne SHG contributions remain zero ($\Delta I_{SH,Het} \sim E_{J_{P1,\pm q_x}} \sim \langle J_{P1,\pm q_x}\rangle = 0$).

These three-mode and four-mode mechanism scenarios are indistinguishable in a one-dimensional single-pump probe experiment. To address this ambiguity, two strategies can be envisaged. A first approach would require the development of an intensity and time-delay-independent local oscillator, which is overlapped on the detector with the second harmonic generated at the sample. By changing the strength of the local oscillator, one could tease out the heterodyne signal, which depends on the field strength of the local oscillator, from the homodyne signal, which is independent of its strength. In this way, one could distinguish three-mode amplification of a pair of low- and high-frequency Josephson plasmons from the four-mode generation of squeezed low-frequency JPPs. However, this is technically challenging as the 400-nm local oscillator should be stabilized in phase down to a small fraction of its 1.3 fs period.

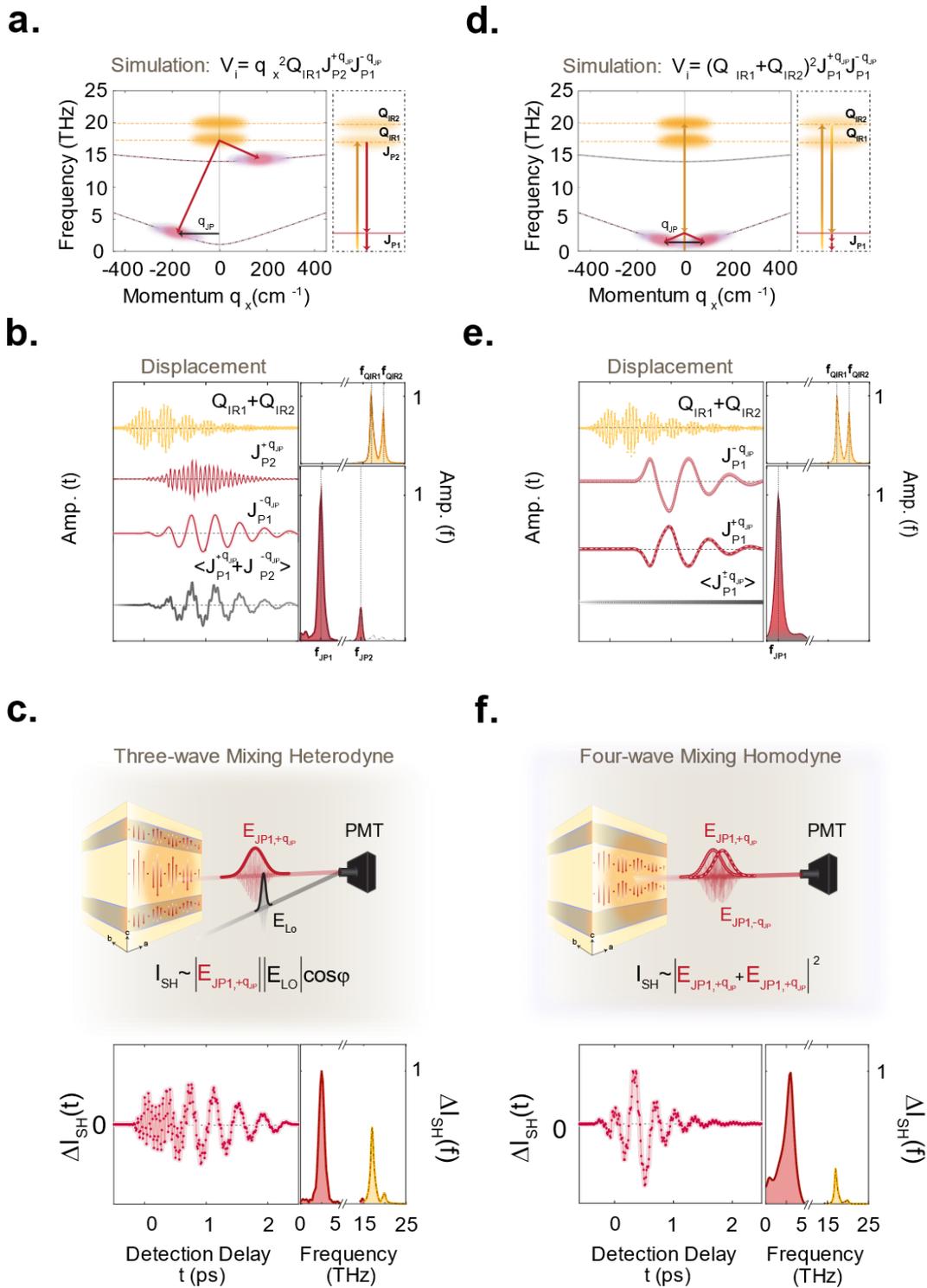

**FIG.2. (a)** Left panel: dispersion curves of the two apical oxygen phonon modes ($Q_{IR1}$ at 17 THz and $Q_{IR2}$ at 20 THz) and of the inter-bilayer ($J_{P1}$) and intra-bilayer ($J_{P2}$) Josephson plasma polaritons along the in-plane momentum $q_x$ are shown as yellow and red dashed lines, respectively. The mid-IR pump excites both apical oxygen phonon modes, which parametrically amplify a pair of inter-bilayer and intra-bilayer Josephson plasma polaritons at finite momentum $q_{JP}$ (black arrow) such that $\omega_{IR1} = \omega_{JP1}(-q_{JP}) + \omega_{JP2}(q_{JP})$. The right panel depicts the energy level diagram corresponding to this model. **(b)** Time-dependent displacement of phonon modes, the two Josephson plasma polaritons and their average value following the photoexcitation, simulated using the three-mode mixing model (left panel) with their respective Fourier spectrum (right panel). The yellow shading indicates the two driven apical oxygen phonons while the two red shading is attributed to Josephson plasma polaritons. **(c)** Second harmonic generation measurement of the Josephson plasmons supercurrents in the heterodyne detection limit. Left panel: the time-delay dependent second harmonic intensity in the heterodyned detection limit following apical oxygen phonon excitation, simulated using the three-mode mixing model detailed in the text, and the corresponding Fourier spectrum using the same color shading as in (b) [8,26]. **(d)** Left panel: same dispersion relations as in (a) now for the four-mode mixing model explained in the text. The mid-IR pump again excites the two apical oxygen phonon modes. Now, they parametrically amplify a pair of inter-bilayer Josephson plasmon polaritons ($J_{P1}$) at finite momentum $\pm q_{JP}$ (illustrated by the black arrow) such that $\omega_{IR2} - \omega_{IR1} = 2\omega_{JP1}(\pm q_{JP})$.. The right panel shows the energy level diagram describing the four-mode mixing model. **(e)** Time-dependent displacement of phonon modes, the two Josephson plasma polaritons and their average value following the photoexcitation, simulated using the four-mode mixing model (left panel) with their respective Fourier spectrum (right panel). With the same color shading as in (b) and (c). **(f)** Second harmonic generation measurement of the Josephson plasmons supercurrents in the homodyne detection limit. From left to right: same as in (c) for four-mode mixing model but in the intermediate detection limit, using the same color shading as in (b), (c) and (e).

---

A second approach, which we explore in this paper, involves the use of higher-dimensional spectroscopy[30-35]. We extend the single-pump second-harmonic probe experiments of Ref.[8] to a two-pump, second-harmonic probe scheme, which is expected to differentiate between three-mode and four-mode scenario. The experimental setup is displayed in Fig. 3a. The two mid-infrared excitation pulses have electric fields denoted by $E_A$ and $E_B$. These two pulses drive the c-axis apical oxygen phonon modes resonantly at two instants in time, separated by a controllable time delay $\tau$. The subsequent coherent dynamics of the JPP and phonon modes are then sampled by a near-infrared probe pulse at a time delay $t$ (defined relative to the arrival time of the last excitation pulse, see

Supplementary Information). The cooperative nonlinear contribution to the tr-SHG intensity $I_{NL}$ from both of the pump pulses can be isolated by subtracting the isolated tr-SHG responses $I_A$ and $I_B$ (to only pulse $E_A$ and $E_B$, respectively), from the response $I_{AB}$ (to both the excitation pulses): $I_{NL} = I_{AB} - I_A - I_B$. Experimentally, we isolate $I_{NL}$ by mechanically chopping the two excitation pulses at frequencies 1/2 and 1/3 of the laser repetition rate $f$ and measuring the tr-SHG intensity component at their difference frequency $f/6$. This procedure is illustrated in Fig. 3b for an excitation pulse delay $\tau$ = 0.5 ps between $E_A$ and $E_B$. The individual tr-SHG signals $I_A$ and $I_B$ each contain a rectified response due to electric-field induced SHG (EFISH), which is then followed by coherent responses of the driven phonons and amplified plasmons. After subtraction, $I_{NL}$ reveals coherent dynamics due to nonlinear terms in the system Hamiltonian. This nonlinear response is also observed to be strongly dependent on temperature, as shown by $I_{NL}$ for two sample temperatures 20 K and 295 K, below and above the critical temperature $T_C$, shown in Fig. 3c.

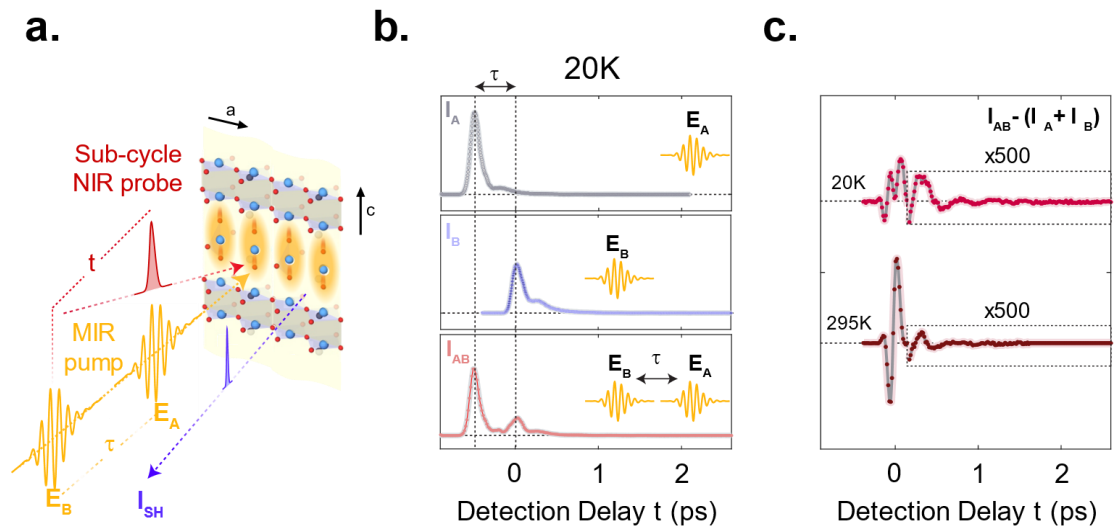

**FIG.3. (a)** Schematic of the same experiment as shown in Fig.1d, now using two CEP stable mid-IR pump pulses $E_A$ and $E_B$ separated by the excitation time delay $\tau$ with approximate fluences of 12 and 6mJ/cm² for $E_A$ and $E_B$, respectively. **(b)** From top to bottom: time-resolved SHG intensities $I_A$, $I_B$ and $I_{AB}$ measured following excitation by only $E_A$, by only $E_B$ (after excitation delay $\tau$) and by both pulses, respectively, at a base

temperature of 20 K (below $T_c$). **(c)** Nonlinear contribution to the time-resolved SHG intensity shown in panel (b), given as $I_{AB} - I_A - I_B$, and for base temperatures of 20K and 295 K (below and above $T_c$, respectively). The dashed rectangle frames the data at later time delays, which are enlarged by a factor of 500 for clarity.

---

Measurements of $I_{NL}$ as a function of delay $\tau$ between two mid-IR pump pulses yielded the two-dimensional time domain maps shown in Figs. 4a and 4b, again for sample temperatures of 20 K and 295 K. For early $\tau$ and t, the rectified component of the homodyne contribution to the nonlinear SHG intensity dominates the response and masks the underlying coherent dynamics. Cropped time-domain data along both time axes, indicated by the black dashed boxes in Figs. 4a and 4d, were used for subsequent Fourier transform (see Supplementary Information). The resulting two-dimensional Fourier spectra, shown in Figs. 4c and 4d for sample temperatures of 20 K and 295 K respectively, each exhibit four dominant peaks. The two peaks at zero detection frequency, $(f_t, f_\tau)$ = (0,17) THz and (0,20) THz, reflect homodyne-detected nonlinear tr-SHG in response to either of the two apical oxygen phonon modes $Q_{IR1}$ and $Q_{IR2}$ (as evidenced by their positions along the vertical $f_\tau$ axis). The positions of the remaining two peaks at (-3;17) THz and (3;20) THz suggest that the dominant ~3 THz response observed in the pump-probe experiment of Fig. 1f (Ref.[8]) is driven cooperatively by excitation of *both* apical oxygen phonon modes.

We next extend the simulations in Fig. 2 to calculate the corresponding multi-dimensional spectra, to identify the coupling mechanism leading to this peak pattern. For the three-wave mixing model in the heterodyne detection, considered in Ref.[8] and Fig. 2b, the resultant two-dimensional spectrum in Fig. 5a shows intense peaks at (-2.5;0) THz, (2.5;0) THz, (-2.5;17) THz and (2.5;17) THz. These peaks arise from the nonlinear coupling between *only* the lower-frequency apical oxygen phonon $Q_{IR1}$ and the two

Josephson plasma polaritons $J_{P1}$ and $J_{P2}$ (see Supplementary Information for details). Clearly, the simulated two-dimensional nonlinear spectrum does not match the measured spectrum, despite the agreement of the one-dimensional spectrum.

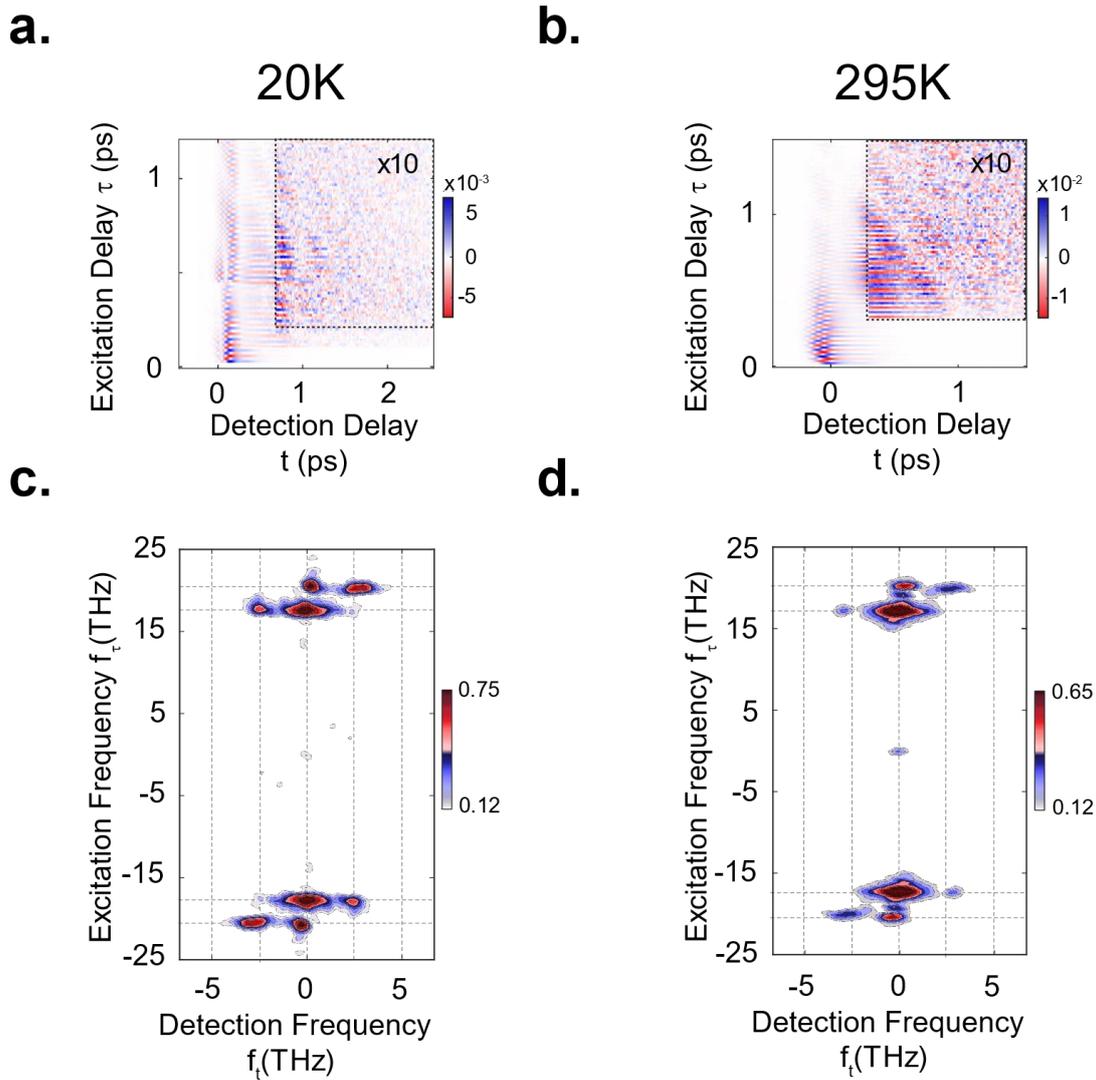

**FIG.4. (a)** Nonlinear contributions to the time-resolved SHG intensity (as described in Fig.3) with the excitation time delay τ changing along the vertical axis, measured at a base temperature of 20 K (below $T_c$). The data inside the black dashed box are multiplied by 10 for clarity. **(b)** Same as in (a) measured at 295 K (above $T_c$). **(c)** Normalized two-dimensional Fourier spectrum of the data inside the black dashed box in panel (a). **(d)** Same as (c) for the data shown in panel (b). Four peaks are found at frequency coordinates (0;17), (0;20), (-3;17) and (3;20), all in units of THz.

For comparison, Fig. 5b shows the two-dimensional spectrum resulting from the four-wave mixing model in the homodyne detection, which also exhibited good agreement with the experiment in the one-dimensional spectrum. Here, we find four dominant peaks at (0; 17) THz, (0;20) THz, (17; -3) THz, and (20; 3) THz. All of these peaks include homodyne contributions of the two driven phonons (as their interference produces a difference-frequency response at 3 THz), and from their cooperative amplification of JPPs. As discussed earlier, the homodyne detection of the amplified JPPs retrieves the non-radiating squeezed state of the amplified Josephson plasmon fluctuations (see Supplementary Information for a detailed discussion). This peak pattern uniquely agrees with the experimental two-dimensional spectrum, thus resolving the ambiguity of the single-pulse pump-probe experiment.

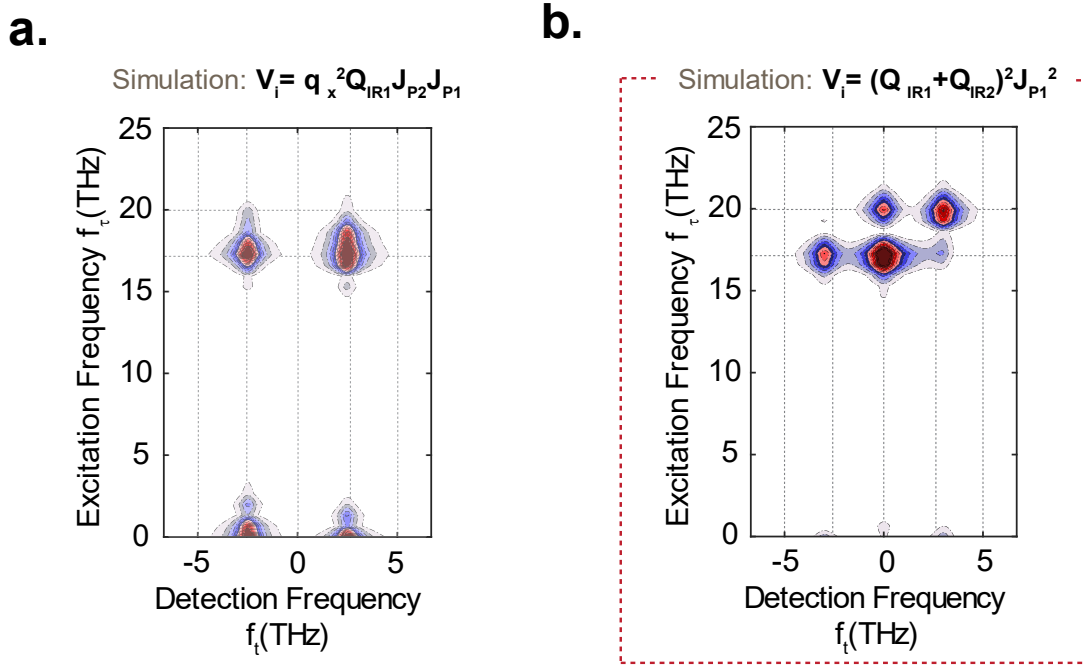

**FIG.5. (a)** Simulated two-dimensional Fourier spectrum for the three-mode mixing model as outlined in Fig.2(b), showing four peaks at corresponding frequency coordinates (-2.5;17), (2.5;17), (-2.5;0) and (2.5;0) all in units of THz. **(b)** Simulated two-dimensional Fourier spectrum for the four-mode mixing model as outlined in Fig.2(d), with four peaks at corresponding frequency coordinates (0;17), (0;20), (-3;17) and (3;20) all in units of THz. The red dashed box emphasizes that the four-mode mixing model is compatible with experimental data shown in Fig.4.

Finally, to confirm the plasmonic origin of the peaks, we display a different set of experiments based on the same multi-dimensional spectroscopy in the higher-doped compound $YBa_2Cu_3O_{6.92}$, in which the apical oxygen phonon frequencies are the same as those of $YBa_2Cu_3O_{6.48}$ [18,36,37], but the zero-momentum inter-bilayer Josephson plasma resonance $\omega_{JP1}$ is blue-shifted to 7 THz [38]. In this situation the four-wave mixing resonance condition is no longer satisfied ($2\omega_{JP1} > \omega_{IR2} - \omega_{IR1}$) such that $J_{P1}$ cannot be effectively amplified. The one- and two-dimensional spectra of this compound should therefore include *only* homodyne mixing between $Q_{IR1}$ and $Q_{IR2}$.

Figure 6a compares the coherent contributions to the single-pump probe tr-SHG intensity for $YBa_2Cu_3O_{6.48}$ ($T_c$ = 48 K) and $YBa_2Cu_3O_{6.92}$ ($T_c$ = 91 K) at a sample temperature of 5 K. Both curves include high-frequency oscillations of the driven phonons, while the lower-frequency oscillations are clearly stronger in $YBa_2Cu_3O_{6.48}$ compared to $YBa_2Cu_3O_{6.92}$. According to the argument above, this should be a result of the additional contribution originating from resonant amplification of the fluctuating inter-bilayer JPPs ($J_{P1,\pm q_x}$). This assumption is confirmed by the corresponding nonlinear two-dimensional spectra shown in Figs. 6b and 6c below the critical temperature $T_C$ (see Supplementary Information for corresponding time-domain data), which exhibit the same signature peak pattern of difference-frequency mixing between the two driven phonons (either in the sample or in the detector), but with different amplitude ratios.

We also observe that the frequency integrated peak amplitudes of the nonlinear two-dimensional spectra exhibit very different temperature dependences. Figure 6d shows that the integrated amplitude in $YBa_2Cu_3O_{6.48}$ has a dominant contribution which decreases with increasing temperature approaching the pseudogap temperature

$T^*$ = 380 K, exhibiting a mean-field behavior proportional to $\sqrt{1 - T/T^*}$ and a temperature-independent contribution (proportional to α) arising from nonlinear phononics for all measured temperatures. In contrast, the frequency integrated peak amplitude in YBa$_2$Cu$_3$O$_{6.92}$ is largely independent of temperature (as shown in Fig. 6e) even up to multiple times $T^*$ = 160 K in this compound. This observation confirms the dominant superconducting origin of the nonlinearities in YBa$_2$Cu$_3$O$_{6.48}$, in contrast to the purely phononic nonlinear response in YBa$_2$Cu$_3$O$_{6.92}$.

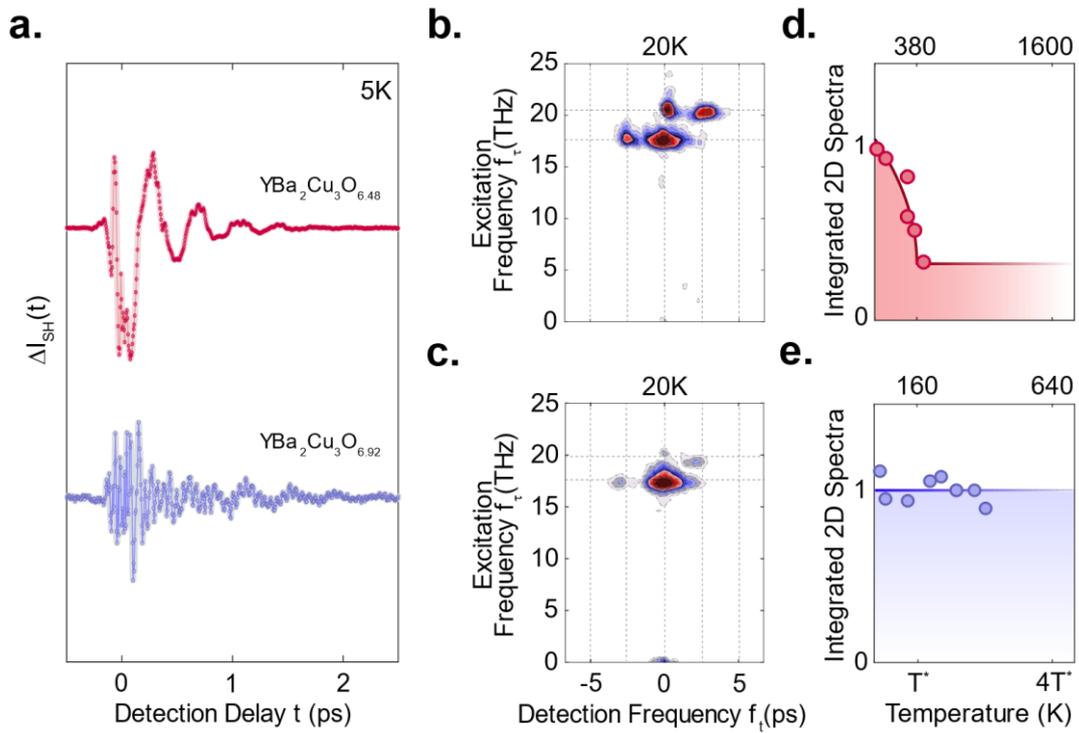

**FIG.6. (a)** Mid-IR pump induced changes in second harmonic intensity as a function of detection time delay t for YBa$_2$Cu$_3$O$_{6.48}$ (red, T$_c$ = 48K) and YBa$_2$Cu$_3$O$_{6.92}$ (blue, T$_c$ = 92K), both at base temperature of 5K (below T$_c$). **(b)** Corresponding normalized nonlinear two-dimensional Fourier spectrum of YBa$_2$Cu$_3$O$_{6.48}$ at base temperature of 20K (below Tc), as in Figure 4(c). **(c)** Same as panel (b) for a different doping of YBa$_2$Cu$_3$O$_{6.92}$. **(d)** Normalized frequency-integrated amplitude of nonlinear two-dimensional spectra of YBa$_2$Cu$_3$O$_{6.48}$ as a function of base temperature (red circles, see Supplementary Information for details). The thick red line is a fit with a mean-field dependence (α + $\sqrt{1 - T/T^*}$), indicating that this quantity has a dominant contribution that dereases as temperature approaches the pseudogap temperature T$^*$ (380K) and a contribution that takes a constant value α for all measured T. **(e)** Same as panel (d) for a different doping of YBa$_2$Cu$_3$O$_{6.92}$. Here, the frequency-integrated nonlinear two-dimensional amplitude does not depend on temperature.

The four-mode mixing model proposed here provides an explanation for observed superconducting-like features in the non-equilibrium THz reflectivity in cuprate compounds. As illustrated in Fig. 7a, excitation of apical oxygen phonon modes at 17 THz and 20 THz leads to coherent amplification of pairs of finite momentum inter-bilayer JPPs which fulfill the resonance condition $2\omega_{JP1}(\pm q_{JP}) = \omega_{IR2} - \omega_{IR1}$ at $\omega_{JP1} \approx 1.5$ THz. These coherently amplified superconducting modes give rise to a characteristic plasma edge at $q = 0$, observed at a frequency blue-shifted relative to its equilibrium value. A Fresnel-Floquet formalism was used to calculate the expected reflectivity of $YBa_2Cu_3O_{6.48}$ under these driven conditions [26,39] (see Supplementary Information for details). The results are shown in Fig. 7b (left panel). Starting from a featureless spectrum, a reflectivity edge emerges near 1.5 THz, in good agreement with experimental data (Fig. 7b (right panel))[2,6,26,39].

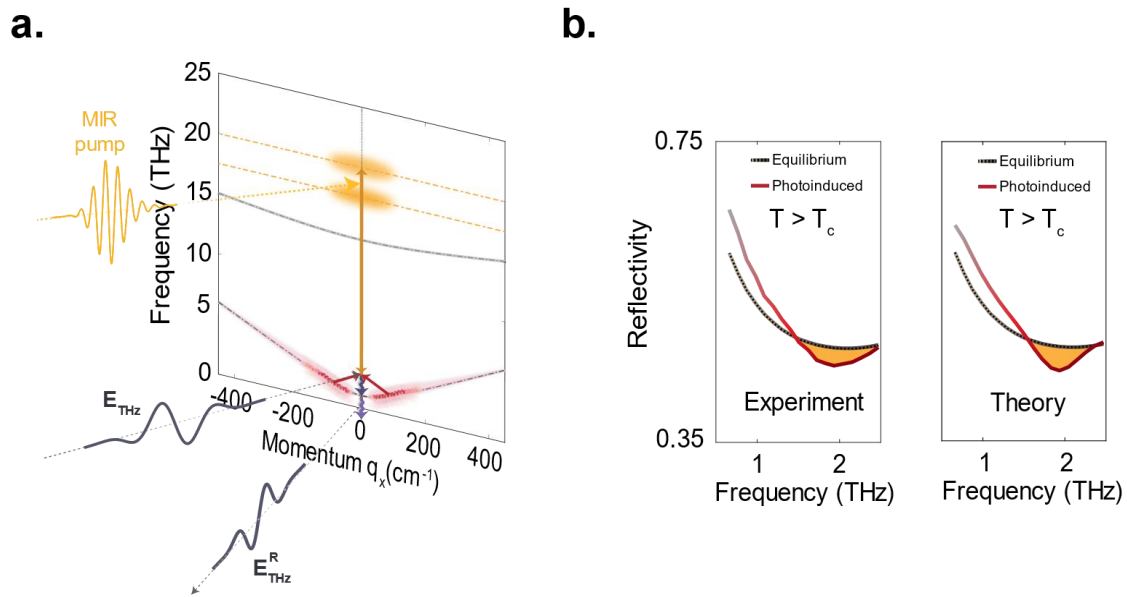

**FIG.7. (a)** Illustration of the coupling between phonon-driven amplified Josephson plasmon polaritons, assuming the four-mode mixing model, and the THz probe field, resulting in the observed photo-induced reflectivity edge. The mid-IR excitation pulse (yellow) resonantly excites the two apical oxygen phonon modes $Q_{IR1}$ and $Q_{IR2}$ (yellow shading) which parametrically amplify pairs of inter-bilayer Josephson plasma polaritons $J_{P1}$ at frequencies $\omega_{JP1}(\pm q_{JP})$ (red shading). These excitations renormalize the reflection

coefficient, as measured by the THz probe field at $q_x = 0$ (grey pulses). **(b)** Comparison between experiment (left,[2]) and theory (right, [26,39]). Dashed black lines show the THz frequency reflectivity above $T_c$ in equilibrium. Red solid lines are the THz frequency reflectivity following mid-IR excitation. In both plots, the yellow shaded area indicates the photo-induced changes at the photo-induced plasma edge ($\omega_{JPLI} < 2$ THz).

---

In summary, multidimensional nonlinear phononic spectroscopy was used to reveal the coupling between apical oxygen phonons and Josephson plasmon polaritons in $YBa_2Cu_3O_{6.48}$. The findings outlined here clarify the ambiguity of the previous one-dimensional measurements and provide support for a four-mode mixing scenario as an explanation for the parametric amplification of Josephson plasmons.

This mechanism implies the generation of squeezed Josephson Plasmons, acting on the dynamics of coherent supercurrents and on their fluctuations. When seen in the context of a phase incoherent superconductor, these squeezed states may point to a mechanism for phase stabilization or phase-noise reduction. This is potentially relevant as a fluctuating pseudogap phase, hosting some form of phase-incoherent superconductivity, is a pre-requisite for the formation of non-equilibrium coherence up to room temperature. Systematic measurements and more extended theoretical work will test these ideas in the future. It also remains to be understood if other materials systems, especially organic superconductors like $K_3C_{60}$[4,9,11] and k-BEDT charge transfer salts[5,10] also undergo some form of parametric amplification of a fluctuating superconducting state at high temperatures. The observation that in all these materials the effect occurs when a strong vortex Nernst effect is seen above $T_c$[40,41] points in this direction. The results reported here suggest a new framework for the engineering of parametrically amplified responses in materials, with potential connections to the physics of time crystals[42-44] and to Floquet quantum matter[45,46].

**Acknowledgements**

We acknowledge C. Trallero-Herrero for illuminating discussions. The research leading to these results received funding from the European Research Council under the European Union's Seventh Framework Program (FP7/2007-2013)/ERC Grant Agreement No. 319286 (QMAC). We appreciate support from the Deutsche Forschungsgemeinschaft (DFG) via the Cluster of Excellence 'CUI: Advanced Imaging of Matter' – EXC 2056 – project ID 390715994 and the priority program SFB925. M. H. Michael and A. Liu received funding from the Alexander von Humboldt Foundation. We thank M. Volkmann and P. Licht for their technical assistance, B. Fiedler and B. Höhling for their support in the fabrication of the electronic devices used on the measurement setup, F. Tellkamp for support relating to the data acquisition software and J. Harms for assistance with graphics.